# Monolithic integration of a high-quality factor lithium niobate microresonator with a free-standing membrane waveguide using femtosecond laser assisted ion beam writing (FLAIBW)

**Zhiwei Fang,**[1,3,4] **Yingxin Xu,**[2] **Min Wang,**[1,3] **Lingling Qiao,**[1] **Jintian Lin,**[1,7] **Wei Fang**[2,8] **and Ya Cheng**[1,5,6,9]

[1] *State Key Laboratory of High Field Laser Physics, Shanghai Institute of Optics and Fine Mechanics, Chinese Academy of Sciences, Shanghai 201800, People's Republic of China*
[2] *State Key Laboratory of Modern Optical Instrumentation, College of Optical Science and Engineering, Zhejiang University, Hangzhou 310027, People's Republic of China*
[3] *University of Chinese Academy of Sciences, Beijing 100049, People's Republic of China*
[4] *School of Physical Science and Technology, ShanghaiTech University, Shanghai 201210, People's Republic of China*
[5] *State Key Laboratory of Precision Spectroscopy, East China Normal University, Shanghai 200062, People's Republic of China*
[6] *Collaborative Innovation Center of Extreme Optics, Shanxi University, Taiyuan, Shanxi 030006, People's Republic of China*
[7] *jintianlin@siom.ac.cn*
[8] *wfang08@zju.edu.cn*
[9] *ya.cheng@siom.ac.cn*

**Abstract:** We demonstrate integrating a high quality factor lithium niobate microdisk resonator with a free-standing membrane waveguide. Our technique is based on femtosecond laser direct writing which produces the pre-structure, followed by focused ion beam milling which reduces the surface roughness of sidewall of the fabricated structure to nanometer scale. Efficient light coupling between the integrated waveguide and microdisk was achieved, and the quality factor of the microresonator was measured as high as $1.67 \times 10^5$.

## References and links

1. A. Guarino, G. Poberaj, D. Rezzonico, R. Degl'Innocenti, and P. Gunter, "Electro-optically tunable microring resonators in lithium niobate," Nat. Photon. **1**(7), 407-410 (2007).
2. T. J. Wang, J. Y. He, C. A. Lee, and H. Niu, "High-quality LiNbO3 microdisk resonators by undercut etching and surface tension reshaping," Opt. Express **20**(27), 28119-28124 (2012).
3. J. T. Lin, Y. X. Xu, Z. W. Fang, J. X. Song, N. W. Wang, L. L. Qiao, W. Fang, and Y. Cheng, "Second harmonic generation in a high-Q lithium niobate microresonator fabricated by femtosecond laser micromachining," arXiv preprint arXiv:1405.6473 (2014).
4. J. T. Lin, Y. X. Xu, Z. W. Fang, M. Wang, J. X. Song, N. W. Wang, L. L. Qiao, W. Fang, and Y. Cheng, "Fabrication of high-Q lithium niobate microresonators using femtosecond laser micromachining," Sci. Rep. **5**, 8072 (2015).
5. C. Wang, M. J. Burek, Z. Lin, H. A. Atikian, V. Venkataraman, I. C. Huang, P. Stark, and M. Lončar, "Integrated high quality factor lithium niobate microdisk resonators," Opt. Express **22**(25), 30924-30933 (2014).
6. J. Wang, F. Bo, S. Wan, W. X. Li, F. Gao, J. J. Li, G. Q. Zhang, and J. J. Xu, "High-Q lithium niobate microdisk resonators on a chip for efficient electro-optic modulation," Opt. Express **23**(18), 23072-23078 (2015).
7. S. Diziain, R. Geiss, M. Steinert, C. Schmidt, W. K. Chang, S. Fasold, D. Füßel, Y. H. Chen, and T. Pertsch, "Self-suspended micro-resonators patterned in Z-cut lithium niobate membranes," Opt. Mater. Express **5**(9), 2081-2089 (2015).
8. W. C. Jiang, and Q. Lin, "Chip-scale cavity optomechanics in lithium niobate," Sci. Rep. **6**, 36920 (2016)

9.  J. T. Lin, Y. X. Xu, J. L. Ni, M. Wang, Z. W. Fang, L. L. Qiao, W. Fang, and Y. Cheng, "Phase-matched second-harmonic generation in an on-chip LiNbO$_3$ microresonator," Phys. Rev. Appl. **6**(1), 014002 (2016).
10. J. Song, J. Lin, J. Tang, Y. Liao, F. He, Z. Wang, L. Qiao, K. Sugioka, and Y. Cheng, "Fabrication of an integrated high-quality-factor (high-Q) optofluidic sensor by femtosecond laser micromachining," Opt. Express **22**(12), 14792–14802 (2014).
11. M. Wang, Y. Xu, Z. Fang, Y. Liao, P. Wang, W. Chu, L. Qiao, J. Lin, W. Fang, and Y. Cheng, "On-chip electro-optic tuning of a lithium niobate microresonator with integrated in-plane microelectrodes," Opt. Express **25**(1), 124-129(2017).
12. Z. Fang, J. Lin, M. Wang, Z. Liu, J. Yao, L. Qiao, and Y. Cheng, "Fabrication of a microresonator-fiber assembly maintaining a high-quality factor by CO$_2$ laser welding," Opt. Express **23**(21), 27941-27946(2015).
13. G. Poberaj, H. Hu, W. Sohler, and P. Günter, "Lithium niobate on insulator (LNOI) for micro-photonic devices," Laser Photon. Rev. **6**(4), 488-503 (2012).
14. L. Cai, Y. Kang, and H. Hu, "Electric-optical property of the proton exchanged phase modulator in single-crystal lithium niobate thin film," Opt. Express **24**(5), 4640-4647(2016).
15. G. H. Shao, Y. H. Bai, G. X. Cui, C. Li, X. B. Qiu, D. Q. Geng, D. Wu, and Y. Q. Lu, "Ferroelectric domain inversion and its stability in lithium niobate thin film on insulator with different thicknesses," AIP Advances **6**(7), 075011(2016).
16. A. Majkic, M. Koechlin, G. Poberaj, and P. Günter, "Optical microring resonators in fluorine implanted lithium niobate," Opt. Express **16**(12), 8769-8779 (2008).
17. D. E. Zelmon, D. L. Small, and D. Jundt, "Infrared corrected Sellmeier coefficients for congruently grown lithium niobate and 5 mol. % magnesium oxide–doped lithium niobate," J. Opt. Soc. Am. B **14**(12), 3319-3322(1997).

## 1. Introduction

Lithium niobate (LN) microresonators have attracted much attention for their broad range of applications in optical signal processing, quantum electrodynamics and optomechanics [1-9]. This is mainly due to the excellent material properties of lithium niobate such as a broad transmission window, large nonlinear optical coefficients, and a large electro-optic tunability. Particularly, the recent advance in the fabrication of high-Q lithium niobate microresonators has promoted the quality factor of such microresonators to $10^5 \sim 10^6$ [3-9]. This is enabled by patterning a lithium niobate thin film bonded onto 2 μm-thickness SiO$_2$ with either focused ion beam [3] or reactive ion dry etching [5]. An additional chemical wet etching in hydrofluoride (HF) acid will selectively remove the SiO$_2$, resulting in a free-standing microdisk that can serve as a whispering-gallery mode (WGM) microresonator. To utilize the high-Q microresonators, light has to be coupled into and out from the microresonator. This is typically achieved using external fiber taper critically coupled to the microdisk. For many applications, it would be desirable to replace the external fiber tapers with monolithically integrated waveguides, thereby the devices can be miniaturized and will be more stable against the disturbance from the environment.

Here, we show that the above-mentioned monolithic integration can be realized with femtosecond laser assisted ion beam writing (FLAIBW), a technique we developed in 2014 for producing high-Q lithium niobate microresonators [3, 4]. The advantages of this technology include the followings. First, both femtosecond laser direct writing and focusing ion beam (FIB) writing are maskless fabrication techniques, which provide the flexibility to achieve rapid prototyping of various kinds of devices. Second, the femtosecond laser is much more efficient than the FIB in terms of the removal rate of material, whilst it is of relatively low fabrication resolution on the order of 1μm as compared to that of FIB. The combination of two approaches allows a high fabrication efficiency mainly determined by the femtosecond laser direct writing and a high fabrication resolution completely determined by the FIB writing. At last, the femtosecond laser direct writing can easily incorporate other functionalities such as microfluidics and microelectrodes into the devices [10-12].

## 2. Sample fabrication and experimental setup

In this work, commercially available ion-sliced Z-cut LiNbO$_3$ (LN) thin film with a thickness of 900 nm (NANOLN, Jinan Jingzheng Electronics Co., Ltd) was chosen for fabricating the on-chip LiNbO$_3$ microdisk resonators. The LN thin film was bonded on a 0.5-mm thick LN substrate sandwiched by a SiO$_2$ layer with a thickness of 2 μm [13-15]. A Ti: sapphire femtosecond laser source (Coherent, Inc., center wavelength: 800 nm, pulse width: 40 fs, repetition rate: 1 kHz) was used for fabricating the on-chip LN microresonator integrated with a waveguide. A variable neutral density filter was used to tune the average power of the laser beam. In the femtosecond laser direct writing, an objective lens (100× / NA 0.80) was used to focus the beam down to a ~1μm-diameter focal spot. The sample could be arbitrarily translated in 3D space at a resolution of 1 μm using a PC-controlled XYZ stage combined with a nano-positioning stage. A charged coupled device (CCD) connecting to the computer was installed above the objective lens to monitor the fabrication process in real time.

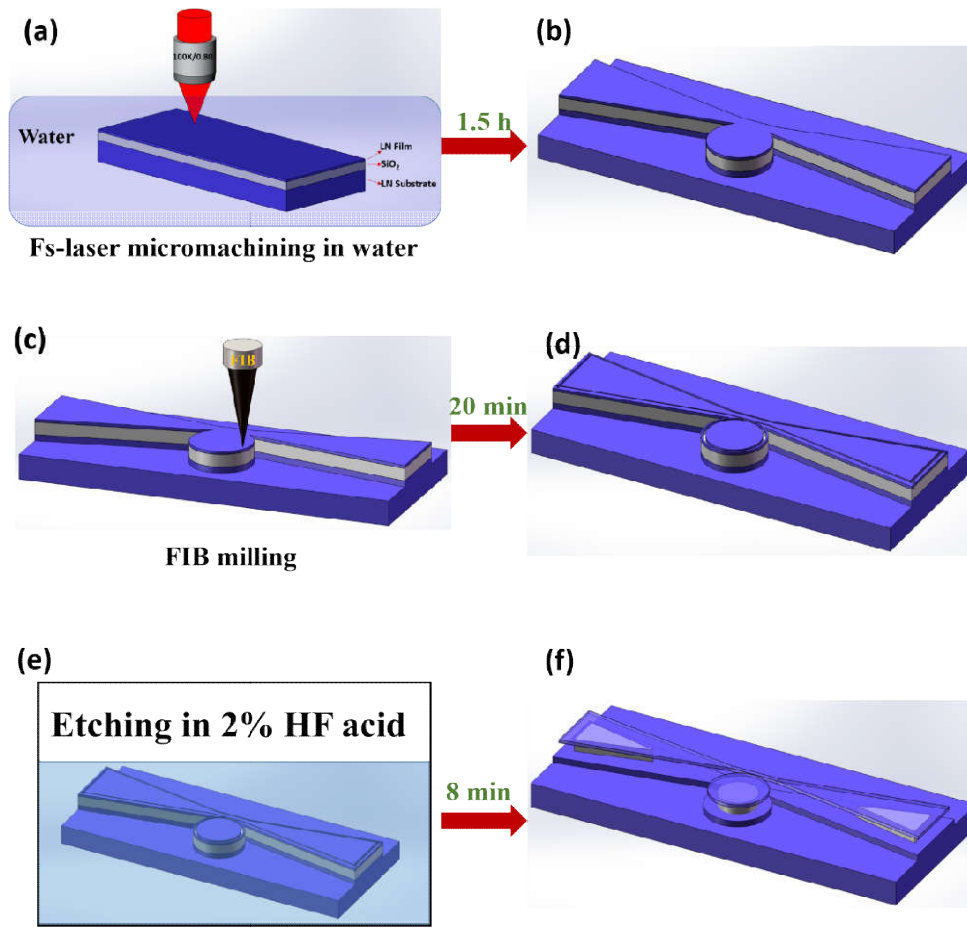

Fig. 1 The processing flow of fabricating an on-chip LN microresonator integrated with a waveguide is illustrated: (a)-(b) Formation of LN microresonator with the integrated waveguide using femtosecond laser microfabrication. (c)-(d) Focused ion beam (FIB) milling to smooth the periphery of the structure fabricated by femtosecond laser direct writing as shown in (b). (e)-(f) Chemical wet etching of the sample undergone the FIB milling to form the freestanding LN microdisk resonator and waveguide.

The procedures of fabricating the integrated device are schematically illustrated in Fig. 1. First, a microresonator-waveguide system was fabricated by ablating the LN substrate immersed in water using tightly focused femtosecond laser pulses, as shown in Fig. 1(a). The height of the patterned microstructure is 10 μm, as schematically illustrated in Fig. 1(b). Next, the periphery of the LN microdisk and waveguide were smoothed using focused ion beam

(FIB) milling, as illustrated in Fig. 1(c). In this step, the microdisk was separated from the waveguide by FIB, as shown in Fig. 1(d). Finally, chemical wet etching, which selectively removes the SiO$_2$ underneath the LN thin film to form free-standing LN microdisk resonator and membrane waveguide, was performed by immersing the fabricated structure in a solution of 2% hydrofluoric (HF) for 8 minutes, as shown in Fig. 1(e). The SiO$_2$ layer was partially preserved to support the LN microdisk and waveguide. The diameter of the LN microdisk is 20 μm. More details of the process flow of fabricating the LN microresonator can be found in [3] and [4].

The LN microdisk was tested using a narrow-band continuous-wave tunable diode laser (New Focus, Model 688-LN). The output laser from a single-mode fiber (SMF-28) was end-fire coupled into the waveguide using a 40× (NA 0.60) microscope objective, and the transmitted light was collected by a 100× (NA 0.80) microscope objective. By using an online fiber polarization controller, WGM modes in the microdisk with certain polarization were excited. A transient optical power detector (Lafayette, Model 4650) was used to measure the transmission spectra.

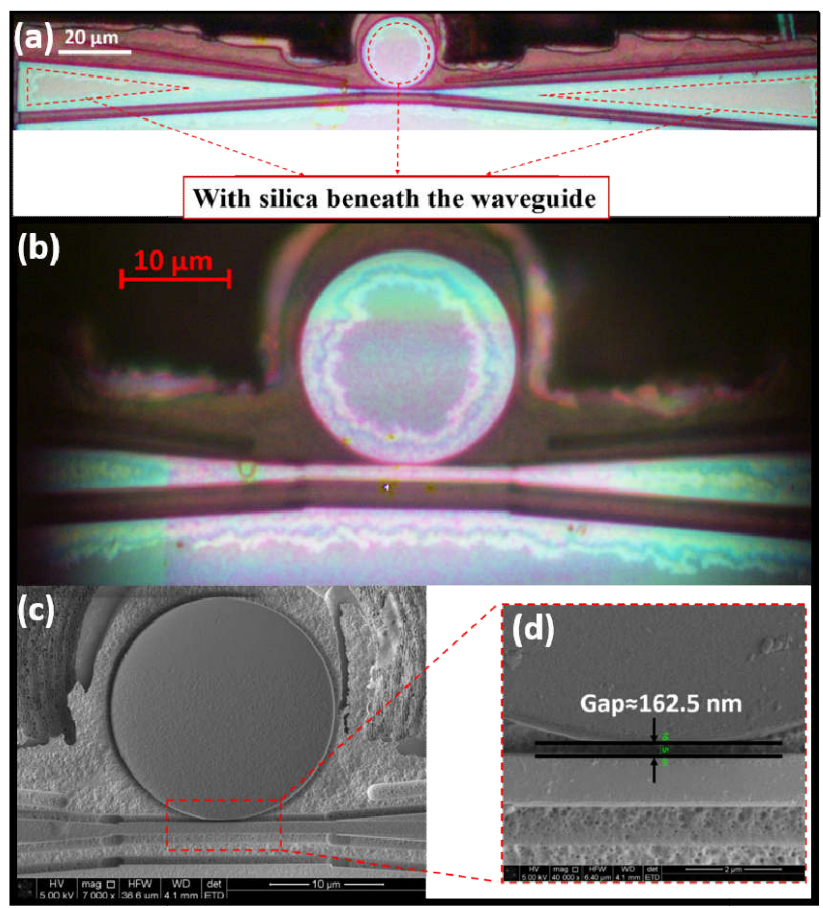

Fig. 2 (a) Top view optical micrograph of the entire integrated device. (b) Close-up view optical micrograph of the microdisk coupled with the waveguide. (c) SEM image of the LN microdisk coupled to the LN waveguide and (d) closed-up view of the coupling area with the gap between the LN microdisk and waveguide of 162.5nm width.

## 3. Results and discussion

Figure 2(a) shows the overview optical micrograph of the fabricated microresonator-waveguide system. It can be seen that a thin waveguide has been fabricated near the microresonator. The waveguide is tapered at its two ends for efficient light coupling into and out of waveguide. For the middle part of the waveguide which is coupled with the microresonator, the width is approximately 1 μm. The areas enclosed in the dashed-line triangles are supported by the $SiO_2$ which survived from the selective chemical etching, thus the waveguide can be fixed on the substrate. A close-up view micrograph shown in Fig. 2(b) provides more details of the fabricated structure, showing the smooth edges of the waveguide and microresonator as a result of FIB milling. Furthermore, the scanning electron micrograph (SEM) in Fig. 2(c) shows that there is a small gap between the microdisk resonator and waveguide. The width of the gap is measured to be 162.5 nm, as indicated in Fig. 2(d). Such a narrow gap is important for achieving the near critical coupling condition which is necessary for obtaining the high-Q factors.

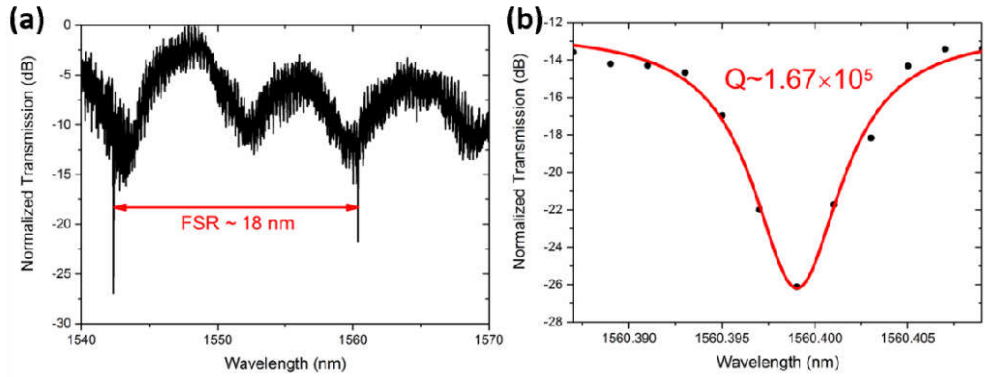

Fig. 3 (a) Normalized transmission spectrum of the integrated device. (b) The Lorentzian fitting shows a Q-factor of $1.67 \times 10^5$.

Figure 3(a) shows the measured transmission spectrum of the integrated microresonator-waveguide system. Based on a Lorentzian fitting in Fig. 3(b), the Q factor was determined to be $1.67 \times 10^5$. It is noteworthy that the Q factor of the integrated LN microdisk-waveguide system is approximately two orders of magnitude higher than the Q-factor previously demonstrated with the integrated LN microring-waveguide system [1, 16].

We further performed the theoretical analysis using a Finite Difference Eigenmode solver (MODE Solutions, Lumerical), which offers both high precision and acceptable computing time. Figure 4(a) presents the model structure used in the simulation, of which all the parameters are the same as that in the experiment. In the model structure, the lithium niobate micordisk has a thickness of 900nm and a diameter of 20 μm. The refractive index of lithium niobate is chosen with a Z cut birefringence [17], which is also consistent with the experimental situation. The waveguide coupled to the microdisk has a width of 1 μm and a length of 20μm. In our simulations, both TE and TM beams with fundamental spatial modes propagated from one end of the waveguide, and the transmission spectra were monitored from the other end of the waveguide after the beams in the waveguide had interacted with the high-Q microresonator. It was found that the calculated transmission spectrum for the TE beam, as shown in Fig. 4(b), can faithfully reproduce the measured result in Fig. 3(a). The free spectrum range (FSR) determined by the spectrum in Fig. 4(b) is 17 nm, which is very close to the measured FSR of 18 nm as indicated in Fig. 3(a). The small difference could be attributed to the fact that it is difficult to precisely determine all the parameters of the microdisk and the waveguide, such as the thickness and the transverse dimensions from an experimental point of view.

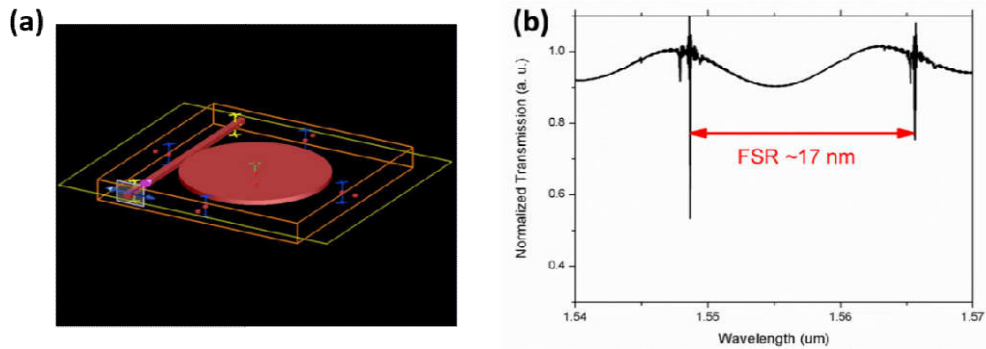

Fig. 4 (a) Illustration of the model structure used in the simulation. (b) Normalized transmission spectrum of the integrated device for an input TE beam.

## 4. Conclusion

To conclude, we have demonstrated an approach for monolithically integrating an optical microresonator with an optical waveguide on lithium niobate substrate. Our technique combines femtosecond laser direct writing and FIB writing, which offers advantages including high fabrication efficiency (compared with that of FIB writing) and high fabrication resolution. The fabricated device shows a quality factor above $10^5$. The technique can also be extended for coupling multiple remotely distributed microresonators using the bus waveguide. The integrated devices are attractive particularly from an application point of view because of the enhanced compactness and stability.

**Funding**